\documentclass[sigconf]{acmart}
\usepackage{balance}
\def\BibTeX{{\rm B\kern-.05em{\sc i\kern-.025em b}\kern-.08emT\kern-.1667em\lower.7ex\hbox{E}\kern-.125emX}}
    
\copyrightyear{2021}
\acmYear{2021}
\setcopyright{acmcopyright}
\acmConference[FPGA '21]{Proceedings of the 2021 ACM/SIGDA International Symposium on Field-Programmable Gate Arrays}{February 28-March 21, 2021}{Virtual}
\acmBooktitle{Proceedings of the 2021 ACM/SIGDA International Symposium on Field-Programmable Gate Arrays (FPGA '21), February 28-March 21, 2021, Virtual}
\acmDOI{10.1145/3431920.3439477}
\acmISBN{978-1-4503-8218-2/21/02}
\usepackage{amssymb}
\usepackage{algorithm}
\usepackage{algpseudocode}
\usepackage{amsmath}
\usepackage{balance}
\usepackage{flushend}
\usepackage{amssymb}
\usepackage{mathtools}

\usepackage{graphicx}
\usepackage{lipsum}
\usepackage{threeparttable}
\usepackage{multirow}
\usepackage{subfigure}
\graphicspath{{fig/}}
\usepackage{booktabs}
\usepackage{chngpage}
\usepackage{mathtools}
\usepackage{verbatim}
\usepackage{amsfonts}
\usepackage{caption}

\DeclareMathOperator\erf{erf}

\setcopyright{none}
\begin{document}

\def\barr{\begin{tabular}{@{}l@{}}}
\def\earr{\end{tabular}}

\fancyhead{}
\title{
NPE: An FPGA-based Overlay Processor for Natural Language Processing
}


\author{Hamza Khan}
\affiliation{%
  \institution{UC Los Angeles}
%  \streetaddress{P.O. Box 1212}
%  \city{Los Angeles}
%  \state{CA}
}

\author{Asma Khan}
\affiliation{%
  \institution{UC Irvine}
%  \streetaddress{P.O. Box 1212}
%  \city{Los Angeles}
%  \state{CA}
}

\author{Zainab Khan}
\affiliation{%
  \institution{Stanford University}
%  \streetaddress{P.O. Box 1212}
%  \city{Los Angeles}
%  \state{CA}
}

\author{Lun Bin Huang}
\affiliation{%
  \institution{Independent Researcher}
%  \streetaddress{P.O. Box 1212}
%  \city{Los Angeles}
%  \state{CA}
}

\author{Kun Wang}
\affiliation{%
  \institution{UC Los Angeles}
%  \streetaddress{P.O. Box 1212}
%  \city{Los Angeles}
%  \state{CA}
}

\author{Lei He}
\affiliation{%
  \institution{UC Los Angeles}
%  \streetaddress{P.O. Box 1212}
%  \city{Los Angeles}
%  \state{CA}
}

%\author{
%Hamza Khan (1), Asma Khan (2), Zainab Khan (3), Lun Bin Huang, Kun Wang (1), %Lei He (1)
%}
%\affiliation{
%\institution{(1) UC Los Angeles, (2) UC Irvine, (3) Stanford University}
%}


%\renewcommand{\shortauthors}{K. Hamza, et al.}


\begin{abstract}

In recent years, transformer-based models have shown state-of-the-art results for Natural Language Processing (NLP). In particular, the introduction of the BERT language model brought with it breakthroughs in tasks such as question answering and natural language inference, advancing applications that allow humans to interact naturally with embedded devices. FPGA-based overlay processors have been shown as effective solutions for edge image and video processing applications, which mostly rely on low precision linear matrix operations. In contrast, transformer-based NLP techniques employ a variety of higher precision nonlinear operations with significantly higher frequency. We present NPE, an FPGA-based overlay processor that can efficiently execute a variety of NLP models. NPE offers software-like programmability to the end user and, unlike FPGA designs that implement specialized accelerators for each nonlinear function, can be upgraded for future NLP models without requiring reconfiguration. We demonstrate that NPE can meet real-time conversational AI latency targets for the BERT language model with $4\times$ lower power than CPUs and $6\times$ lower power than GPUs. We also show NPE uses $3\times$ fewer FPGA resources relative to comparable BERT network-specific accelerators in the literature. NPE provides a cost-effective and power-efficient FPGA-based solution for Natural Language Processing at the edge.

\end{abstract}

%\begin{CCSXML}
%<ccs2012>
%  <concept_id>10010520.10010553.10010562</concept_id>
%  <concept_desc>Computer systems organization~Embedded systems</concept_desc>
%  <concept_significance>500</concept_significance>
%  <concept_id>10010520.10010575.10010755</concept_id>
%  <concept_desc>Computer systems organization~Redundancy</concept_desc>
%  <concept_significance>300</concept_significance>
%  <concept_id>10010520.10010553.10010554</concept_id>
%  <concept_desc>Computer systems organization~Robotics</concept_desc>
%  <concept_significance>100</concept_significance>
%  <concept_id>10003033.10003083.10003095</concept_id>
%  <concept_desc>Networks~Network reliability</concept_desc>
%  <concept_significance>100</concept_significance>
%</ccs2012>
%\end{CCSXML}
%
%\ccsdesc[500]{Computer systems organization~Embedded systems}


\maketitle
%  \caption{Seattle Mariners at Spring Training, %2010.}
%  \Description{Enjoying the baseball game from %the third-base


\section{Introduction}

One of the most common Natural Language Processing (NLP) tasks is sequence transduction, translating an input sequence to an output sequence. Traditionally, convolutional neural networks (CNNs) and recurrent neural networks (RNNs) have been used for this task \cite{graves2012sequence, gehring2017convolutional}. The Transformer architecture \cite{vaswani2017attention} removes all recurrent and convolutional components and instead relies on self-attention mechanisms, typically showing better computation time and performance. A transformer is made up of two parts, encoders and decoders. The Bidirectional Encoder Representations from Transformers (BERT) model \cite{devlin2018bert} incorporates the encoders from transformers to generate a state-of-the-art language representation model. When it was first introduced, BERT broke records on eleven different NLP tasks. Since then, variations of BERT such as RoBERTa \cite{liu2019roberta} and DistilBERT \cite{sanh2019distilbert} have shown even better performance and accuracy. Implementing efficient accelerators for inference of these transformer models has proven a challenging task.

FPGA-based overlay processors have provided effective solutions for CNN and other network inference on the edge, allowing for flexibility across networks without reconfiguring the FPGA \cite{yu2019opu, yu2020light, yu2020uni}. CNNs consist mostly of linear matrix operations like convolution and pooling, and have consistently shown resilience to extreme quantization. However, BERT and its successors \cite{liu2019roberta, sanh2019distilbert, jiao2019tinybert, lan2019albert} cannot be efficiently accelerated using existing CNN FPGA accelerators. Although they compute many matrix multiplications, transformer models also introduce complex nonlinear operations that require higher precision, and are called with higher frequency. For instance, softmax and layer normalization \cite{ba2016layer} are performed several times in each encoder for BERT, and block subsequent computation until they are finished processing. As such, it is essential to calculate them efficiently while maintaining high throughput. These nonlinearities must be computed on-device for performance-sensitive applications, as sending to a CPU causes significant latency overhead and is not practical on the edge.

Most existing accelerators \cite{mlsys2020_143, ham20203} include specialized units for computing each type of nonlinearity. For instance, FTRANS \cite{li2020ftrans}, the only previously published FPGA accelerator for transformers, includes separate softmax and layer normalization modules. Since NLP is a constantly-evolving field that may introduce different types of nonlinearities, this specialized approach means that an FPGA design may need reconfiguration for additional NLP networks. It also leads to unnecessary area overhead and under-utilized resources across nonlinear operations.

\begin{figure}[t]
	\includegraphics[width=1\columnwidth]{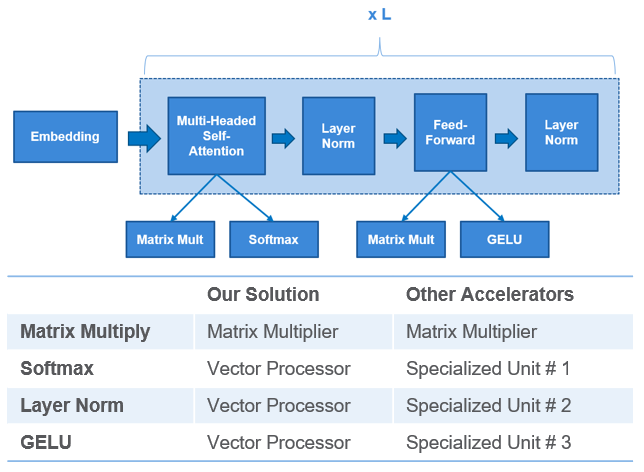}
	\centering
	\caption{BERT network structure, showing how each operation would be mapped onto NPE compared to a conventional network-specialized accelerator.}
	\label{bert_architecture}
\end{figure}

% TODO: change this caption? Maybe add something like "NPE has better resource sharing and lower area overhead"? Idk...

In this paper we propose NPE, an FPGA-based overlay processor for NLP model inference at the edge. As shown in Figure \ref{bert_architecture}, unlike most other accelerators, NPE employs a common method for approximating different nonlinear functions efficiently and accurately without added overhead. The main contributions of our work are as follows: 

\begin{itemize}
    \item We design a software-programmable domain-specific overlay processor with a matrix multiply unit and a multi-precision vector unit for NLP processing.
    \item We employ a unified piecewise polynomial approach for nonlinear function approximation to allow extensibility to future nonlinear functions that may be required.
    \item We demonstrate that our proposed accelerator can meet the real-time latency constraints for conversational AI while maintaining $4\times$ and $6\times$ lower power than GPUs and CPUs respectively. Our design utilizes $3\times$ fewer FPGA resources relative to a comparable network-specific transformer FPGA accelerator.
\end{itemize}

\section{Related Work}\label{Related Work}

While BERT has been highly optimized at the software level for CPU and GPU, little work has been published related to custom hardware acceleration of any transformer-based networks, particularly on FPGAs. Two ASICs have been recently proposed, \textit{OPTIMUS} \cite{mlsys2020_143} and \textit{$A^3$} \cite{ham20203}, that each accelerate different parts of transformer inference. \textit{OPTIMUS} optimizes matrix multiplication for transformers by exploiting sparsity. It has dedicated exponential, divider, and square root components for nonlinearities, leading to wasted area since each is only used a small fraction of the time. \textit{$A^3$} accelerates attention mechanisms using various approximation methods. It has a very deep pipeline that is specialized for attention, which is inefficient from an overall design point of view because the multipliers and function approximation units cannot be reused even for other matrix multiplications. In particular, $A^3$ would need to be paired with a conventional matrix multiply accelerator to implement BERT. At any one time, either the $A^3$ unit or the main accelerator would be computing, leading to many idle resources and wasted performance potential.

To the best of our knowledge, \textit{FTRANS} \cite{li2020ftrans} is the only currently published FPGA accelerator for BERT and related transformer networks. FTRANS takes a very specialized approach for implementing transformers, in which it has dedicated encoder and decoder modules. Each of these modules has many specialized components. For example, the attention head module contains components for softmax and layer normalization, as well as five unique PE banks to perform each matrix multiply subroutine required (see Table \ref{bert_compute} for the computation in an attention layer). While it is sufficient to implement transformers, FTRANS is not flexible enough to handle any non-transformer network. As the NLP state-of-the-art evolves and new network variants emerge, the FTRANS architecture may have to be extensively redesigned to adapt.

\section{Background}\label{Background}

\subsection{Conversational AI}

Low power and low latency NLP is a prerequisite for conversational AI at the network edge. Conversational AI allows people to interact naturally with machines in a dialogue. For instance, a user may ask a question to a smart speaker and expect a human-like response almost immediately. For this response to seem natural, it must be returned within 300 milliseconds. In these 300 ms, the device must perform several complex steps including speech-to-text, user authentication, and natural language processing. As a result, any single model's inference should be complete within 10-15 milliseconds. Recently, many efforts have been made by the GPU community to optimize GPU implementations of BERT to reach this threshold.

\subsection{The BERT Model}

BERT adopts the structure of the encoders from transformers. While there are many BERT variants, the particular structure can be described by three parameters: number of encoders $L$, number of attention heads $A$, and hidden layer size $H$. We focus on $\text{BERT}_{\text{BASE}}$, which is composed of 12 encoders, each with 12 attention heads and a hidden size of 768 ($L$ = 12, $A$ = 12, $H$ = 768). Figure \ref{bert_architecture} shows the general structure for BERT.

The model starts with an embedding layer that converts each input language sequence into features. For instance, an input sequence with 512 tokens in $\text{BERT}_{\text{BASE}}$ would be converted to a $512 \times 768$ matrix, where each token is replaced by a 768-length feature vector. Here, a token refers to a few adjacent characters, where a word is made up of one or more tokens. The embedding step has negligible computation but requires lots of memory. Therefore, we assume this initial step is performed off-chip and focus on accelerating the computationally-intensive encoders.

Embedding is followed by $L$ = 12 encoders. Each encoder performs four back-to-back operations: multi-headed self-attention, layer normalization, feed-forward layers, then layer normalization again. The encoder calculation can be decomposed into matrix-matrix multiplies followed by one of three primitives: softmax, layer normalization, and the GELU activation \cite{hendrycks2016gaussian}. Table \ref{bert_compute} describes the computation in further detail.

\begin{table}
  \caption{Computation for BERT with \textit{A} attention heads. $X$ represents the input to an encoder. $X_1$-$X_5$ represent subsequent outputs after each operation. The computation is broken down into matrix-matrix multiplies, softmax, layer normalization, and GELU.}
  \label{bert_compute}
  \begin{tabular}{l}
			\toprule
			\textbf{Embedding}\\
			\midrule
			$X$ = Embedding(input\_sequence) \\
			\midrule
			\textbf{Multi-Headed Self-Attention} \\
			\midrule
			for (i = 0 to A - 1) \\
			\hspace{12pt} $Q_i$ = $X$ $W_{Q_i}$; \hspace{4pt} $K_i$ = $X$ $W_{K_i}$; \hspace{4pt} $V_i$ = $X$ $W_{V_i}$ \\
			\hspace{12pt} $Z_i$ = softmax($\frac{Q_i{K_i}^T}{k}$)$V_i$ \hspace{8pt} where $k$ is a known constant \\
			$X_1$ = [$Z_0$, ..., $Z_{A-1}$] $W_O$ \\
			\midrule
		    \textbf{Layer Normalization A} \\
			\midrule
			$X_2$ = LayerNorm($X$ + $X_1$) \\
			\midrule
			\textbf{Feed-Forward} \\
			\midrule
			$X_3$ = GELU($X_1$ $W_1$ + $b_1$) \\
			$X_4$ = $X_3$ $W_2$ + $b_2$ \\
			\midrule
		    \textbf{Layer Normalization B} \\
			\midrule
			$X_5$ = LayerNorm($X_2$ + $X_4$) \\
			\bottomrule
\end{tabular}
\end{table}

The BERT model is typically trained to handle sequences of up to 512 tokens. For any particular task we pick a sequence length $seq\_len \leq 512$, padding any input sequences less than $seq\_len$ and truncating any sequences larger than $seq\_len$. This parameter determines the complexity of the operations performed and the inference speed. Sequence lengths less than 32 are usually too small for practical applications. Typical benchmarks use sequence lengths of 64 or 128.

\subsection{BERT Nonlinear Operations}

\subsubsection{\textbf{GELU Activation}}

The GELU activation is defined by the following equation:

\begin{equation}
  \text{GELU}(x) = xP(X \leq x) = x \cdot \frac{1}{2}[1+\erf{(x/\sqrt{2})}]
\end{equation}

It is commonly approximated using the $\tanh$ function as in Equation \ref{eqn:gelu} and can also be approximated directly using a lookup table.

\begin{equation}
    \label{eqn:gelu}
    \text{GELU}(x) \approx 0.5x(1+\tanh{[\sqrt{2/\pi}(x+0.044715x^3)]})
\end{equation}

\subsubsection{\textbf{Layer Normalization}}

Layer normalization first requires computing the mean and variance of a matrix across rows. Given a matrix $x$ of dimension $N \times K$, we compute the mean and variance for row $i$.

\begin{equation}
    \label{sigma_mu}
    \mu_i = \frac{1}{K} \sum_{k=1}^{K} x_{i,k}; \hspace{12pt} \sigma_i^2 = \frac{1}{K} \sum_{k=1}^{K} (x_{i,k} - \mu_i)^2
\end{equation}

Then, the mean and variance are applied to each element using Equation \ref{norm_eqn} to get the normalized output $\hat{x}$. Finally, each $\hat{x}_i$ is scaled and shifted by trained vectors $\gamma$ and $\beta$ to get the layer normalization output $y$, as shown in Equation \ref{lnorm_eqn}.

\begin{equation}
    \label{norm_eqn}
    \hat{x}_{i,k} = \frac{x_{i,k}-\mu_k}{\sqrt{\sigma_k^2+\epsilon}}
\end{equation}

\begin{equation}
    \label{lnorm_eqn}
    y_{i,k} = \hat{x}_{i,k} \cdot \gamma_k + \beta_k
\end{equation}

\subsubsection{\textbf{Softmax}}

The definition of softmax is shown in Equation \ref{eqn_softmax}. Softmax can be difficult to implement in hardware because of the exponential and division operations. It can be directly realized using dedicated exponential and division units, at the cost of under-utilized resources and extra area overhead.

\begin{equation}
    \label{eqn_softmax}
    \text{softmax}(x_j) = \frac{e^{x_j}}{\sum_{i}e^{x_i}}
\end{equation}

\subsection{Throughput Requirements of Nonlinear Operations}

Since matrix multiply operations depend on the results from preceding nonlinear operations, nonlinear processing needs to have high enough throughput to not add significant latency overhead. The throughput requirement for nonlinear operations can be determined by the number of elements we need to process and the cycles available for processing. We define the cycle budget for a nonlinear operation as the number of cycles that the preceding matrix multiply takes to process, given the matrix multiply dimensions and the number of multiplies per cycle. For $\text{BERT}_{\text{BASE}}$, given 2048 multiplies per cycle and a sequence length of 512, we show the throughput requirements for each nonlinearity in Table \ref{throughput}. The layer normalization after attention and the one after GELU are shown separately, since they have different throughput requirements.

\begin{table}[!thb]
	\centering
	\caption{Throughput requirements (elements per cycle) for different nonlinearities in $\text{BERT}_{\text{BASE}}$ with sequence length of 512 and 2048 multiplies per cycle. The matrix to be processed has dimensions $N \times M$.}
	\resizebox{0.5\textwidth}{!}{%
		\begin{tabular}{@{}lllllll@{}}
			\toprule
			Nonlinearity & N & M & Cycle Budget & Throughput & \% of Overall Cycles \\
			\midrule
			Softmax & 512 & 512 & 8,192 & 32 & 5\\
			Layer Norm A & 512 & 768 & 147,456 & 2.7 & 7.5 \\
			GELU & 512 & 3072 & 589,824 & 2.7 & 30 \\
			Layer Norm B & 512 & 768 & 589,824 & 0.7 & 30 \\
			\bottomrule
		\end{tabular}
	}
	\label{throughput}
\end{table}

The final column of Table \ref{throughput} indicates percentage of overall cycles that depend on each nonlinear computation. Specifically, this value shows the percentage of overall matrix multiply cycles that are followed by the nonlinearity. For instance, we see that 30\% of the overall cycle time is spent computing the matrix multiply inputs to GELU operations.

From Table \ref{throughput}, we see that Layer Normalization and GELU both require a throughput average of less than three elements per cycle,  while softmax requires 32 elements per cycle to throughput-match the matrix multiply operations. To put this in perspective, this means that without additional optimizations we would need to perform softmax on a vector of 512 elements in just 16 cycles.

\section{Nonlinearity Processing}\label{Nonlinearity Processing}

\subsection{Nonlinear Function Approximation}

There are dozens of ways to approximate nonlinear functions. We first discuss several specialized approaches that are commonly used for different nonlinear functions. We then discuss our efficient uniform approach to approximate several types of nonlinearities.

\subsubsection{\textbf{Specialized Approaches}}

Approximating complex nonlinear functions has been a field of interest for many decades. Common functions like softmax and square roots have been implemented in many different ways over the years using mathematical models with varying computational complexities and accuracies. The square root function, for example, can be approximated using a Taylor Series Expansion \cite{kwon2009floating}, the Chebyshev approximation \cite{sawada2002mechanical}, the Newton-Raphson algorithm \cite{cornea1999correctness}, and the CORDIC algorithm \cite{volder1959cordic, andraka1998survey}. It can also be approximated directly using a lookup table (LUT). Softmax is also often implemented using one or more LUTs for exponential calculation \cite{yuan2016efficient, du2019efficient}.

For reasonable accuracy with the lookup-based approaches, the LUTs typically are large and require a lot of memory. In contrast, piecewise linear approaches have shown good accuracy while maintaining very small lookup tables \cite{frenzen2010number}.

\subsubsection{\textbf{Unified Nonlinearity Processing}}

As previously shown, each type of nonlinearity has several ways it can be approximated. However, taking a separate approach for each function can lead to many times more area and much more underutilized resources. For instance, most transformer accelerators have dedicated exponential units and dividers for softmax and dedicated square root units for layer normalization.

We take a more unified approach that uses only piecewise polynomial approximation along with simple vector operations to process various nonlinearities. Some functions, like GELU, can directly be approximated using piecewise approximation. Others, like softmax, may use piecewise approximation for intermediate functions like  exponentials or square roots and then use adders, multipliers, etc. for the remainder of the computation. To implement this unified approach, our hardware is optimized for piecewise function approximation as well as other operations like multiplication, addition/subtraction, and vector reduction (e.g., max, sum).

\subsubsection{\textbf{Multi-precision Computation}}

Unlike matrix multiplies and other linear operations, nonlinear computation cannot be fully quantized to low-precision data types (such as 8-bit fixed point numbers) while maintaining network accuracy. Instead, different steps of the nonlinear computation require varying levels of precision. For instance, layer normalization may take in a 16-bit fixed point number as an input but will need to process the variance calculations using 32 or even 64-bit fixed point. As such, all supported operations (arithmetic, reduction, etc.) need to be flexible enough to operate on several different data types.

\subsection{Piecewise Function Approximation}

\subsubsection{\textbf{General Piecewise Approximations}}

\begin{figure}[t]
	\includegraphics[width=1\columnwidth]{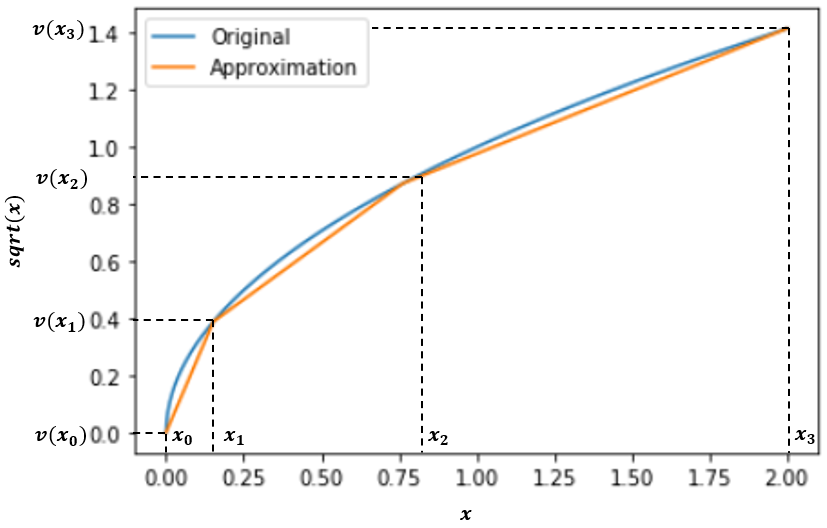}
	\centering
	\caption{Piecewise linear approximation of $\sqrt{x}$ on the interval [0, 2) with three segments.}
	\label{pwlin_approx}
\end{figure}

Piecewise function approximation involves breaking up a nonlinear function of interest into smaller regions along a chosen interval, each approximated by a simpler function such as a line or polynomial. Continuous Piecewise Linear (CPWL) approximations, in particular, use lines for approximation where the end point of each region is the same as the start point of the next region. Figure \ref{pwlin_approx} shows an approximation of $v(x) = \sqrt{x}$ with just three segments. The starting $x$ and $v(x)$ values for each segment are called knot samples and nodal values respectively.

Typically, approximation regions are chosen to be either uniform width or non-uniform width. Uniform width segments tend to be easier to evaluate but require significantly more segments. Non-uniform width segments can better approximate functions with fewer segments but require more complex evaluation. The advantage of non-uniform segments in terms of total number of segments required becomes even more significant when dealing with functions that have large mostly-linear regions. In particular, consider functions like GELU(x) and $\sqrt{x}$ which are nearly linear except for a very small nonlinear region near zero. Uniform segmentation would require orders of magnitude more segments than non-uniform to approximate the linear regions, leading to significantly more memory usage for the same maximum approximation error.

We can also use a piecewise polynomial approach to approximate some functions, which takes more cycles to compute but gives higher accuracy. For most functions, including those needed for BERT, piecewise linear is enough. We discuss our non-uniform continuous piecewise linear segmentation in the following section.

\subsubsection{\textbf{Piecewise Linear Approach}}

\begin{figure*}[t]
	\includegraphics[width=2\columnwidth]{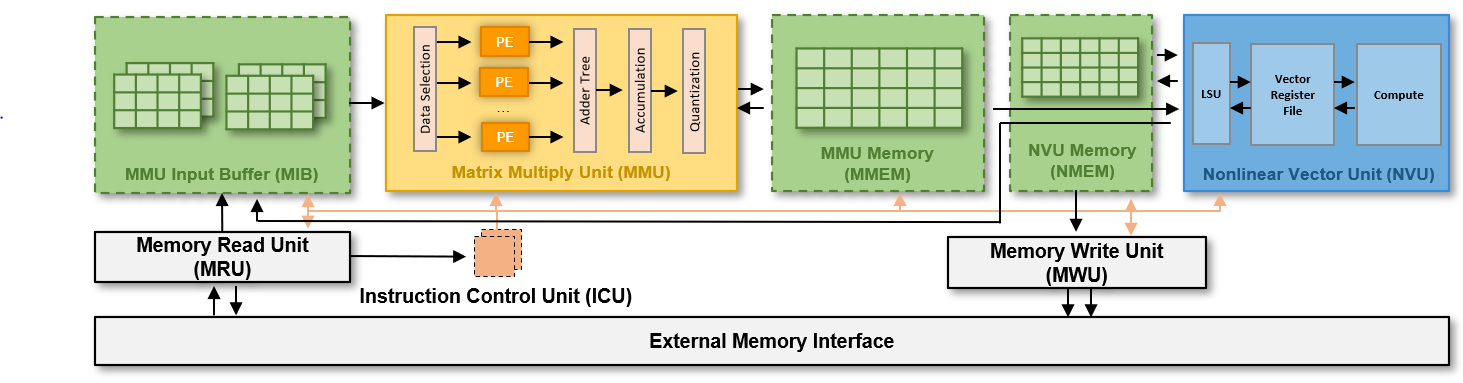}
	\centering
	\caption{Overall architecture of NPE. The memory read unit (MRU) reads from external memory and writes to the MMU input buffer (MIB). The MMU performs matrix multiplication and writes results to its scratchpad memory. The NVU reads from the MMU scratchpad memory and writes intermediate results to the MMU input buffers. The memory write unit (MWU) transfers final results from the NVU scratchpad to external memory.}
	\label{overall_architecture}
\end{figure*}

The main considerations for piecewise linear approximations are the number of segments, segment widths, and the best approximation for each segment. Frenzen et al. \cite{frenzen2010number} explored the number of segments needed for piecewise linear approximation of various common functions using different segmentation techniques. Based on their results, we see that it is possible to maintain high accuracy with very few segments (even less than 10, depending on accuracy constraints). We use a segmentation approach based on \cite{berjon2015optimal}, which describes one method for finding an optimal partition with non-uniform segmentation. We find that even sub-optimal segmentation can result in no accuracy loss for BERT inference on the test set. Algorithm \ref{cpwl_eval} gives the general computation for evaluating a continuous piecewise linear function.

\begin{algorithm}[H]
    \begin{flushleft}
    1) Find the sub-interval containing the input value $x$ given that:
    \end{flushleft}
    $x_{i-1} \leq x < x_i$ \\
    \begin{flushleft}
    2) Find the fractional distance $\delta$ from $x_{i-1}$:
    \end{flushleft}
    $\delta = (x - x_{i-1})/(x_i-x_{i-1})$  \\
    \begin{flushleft}
    3) Determine the final piecewise linear approximation:
    \end{flushleft}
    $v(x) \approx (1-\delta) \hspace{2pt} v(x_{i-1}) + \delta \hspace{2pt} v(x_i)$ \\
    \caption{Continuous Piecewise Linear Approximation of a function $v(x)$ based on \cite{berjon2015optimal}. Here, the knot samples are $x_0$, ..., $x_N$ and the corresponding nodal values are $v(x_0), ..., v(x_N)$.} \label{cpwl_eval}
\end{algorithm}

Finding the sub-interval from Step 1 of Algorithm \ref{cpwl_eval} adds additional complexity. If using uniform segmentation, the segment number can be found simply by using the upper bits of the input. Since we use non-uniform segmentation, we implement more complex segment address calculation like that used in \cite{lee2003non}. In software, this segment address calculation could be performed using Algorithm \ref{seg_addr}. On a CPU or GPU, these calculations would take tens of instructions. Meanwhile, with specialized hardware, this task can easily be done on an FPGA or ASIC within a single clock cycle, such as with a priority encoder.

\begin{algorithm}[H]
    \caption{Software evaluation of non-uniform piecewise linear segment address for input $x$ with knot samples $x_0$, ..., $x_N$.} \label{seg_addr}

    \begin{algorithmic}
        \For{$i\gets N, 0$}
            \If{$x$ $\geq$ $x_i$}
                \State \Return i
            \EndIf
        \EndFor
    \end{algorithmic}

\end{algorithm}

By itself, piecewise linear approximation is not always accurate enough to be used without a large number of segments. However, with normalization and range limiting of the fixed point input and subsequent denormalization of the output, this approximation can maintain high accuracy with only a few segments.

\section{Accelerator Architecture} \label{Overall Architecture}

In this section, we present the overall architecture of NPE, our FPGA overlay processor for NLP. NPE adopts several components from OPU \cite{yu2019opu} including the matrix multiply unit (MMU). Figure \ref{overall_architecture} shows the architecture of the accelerator.

\subsection{Accelerator Modules}

The NPE architecture consists of the following: instruction control unit (ICU), memory read unit (MRU), memory write unit (MWU), matrix multiply unit (MMU), and the nonlinear vector unit (NVU). The ICU sends instructions to each of the modules. Data is exchanged between the functional units through a series of memory buffers.

\subsection{Data Flow}

The memory read unit (MRU) reads data from external memory and writes it to the MMU's input buffer (MIB). The MMU reads from the MIB and uses its scratchpad memory (MMEM) both for intermediate computations and for its final output. The NVU uses its scratchpad memory (NMEM) for its intermediate operations. The NVU accesses MMEM for its input data and deposits the final results either to the MIB (MMU input buffer) or to the NMEM (NVU scratchpad) for subsequent retrieval by the MWU, which writes these results to the external memory. The operations of the MRU, MWU, MMU and the NVU are pipelined to hide off-chip memory latency and to allow for maximum concurrency between these units.

\subsection{Matrix Multiply Unit (MMU)}

The MMU computation consists of five stages: data selection, inner product, adder tree reduction, accumulation, and quantization. Data selection loads the necessary matrix operands from the input buffers and rearranges them as needed. The matrix multiplication is implemented using an array of PEs, where each PE performs an inner product using an array of multipliers followed by an adder tree. Then, the PE outputs can be further summed up by another adder tree, with the number of final outputs dependent on the matrix multiplication dimensions. Finally, the inner product outputs are accumulated and then quantized.

The MMU implementation has 128 PEs with 16 multiply accumulate units each (for a total of 2048 multipliers). These multiply accumulate units map to the FPGA's DSP slices in our implementation. There are two versions of the NPE design, one supporting 8-bit matrix multiplies and the other supporting 16-bit matrix multiplies. The 16-bit version uses each DSP slice for a single element multiply, for a total throughput of 2048 multiplies per cycle. The 8-bit version decomposes each DSP slice into two 8-bit multipliers with one input in common (due to DSP slice resource constraints). On the same board, we can get a throughput of 4096 8-bit multiplies per cycle with 2048 DSP slices.

\begin{figure*}[t]
	\includegraphics[width=1.75\columnwidth]{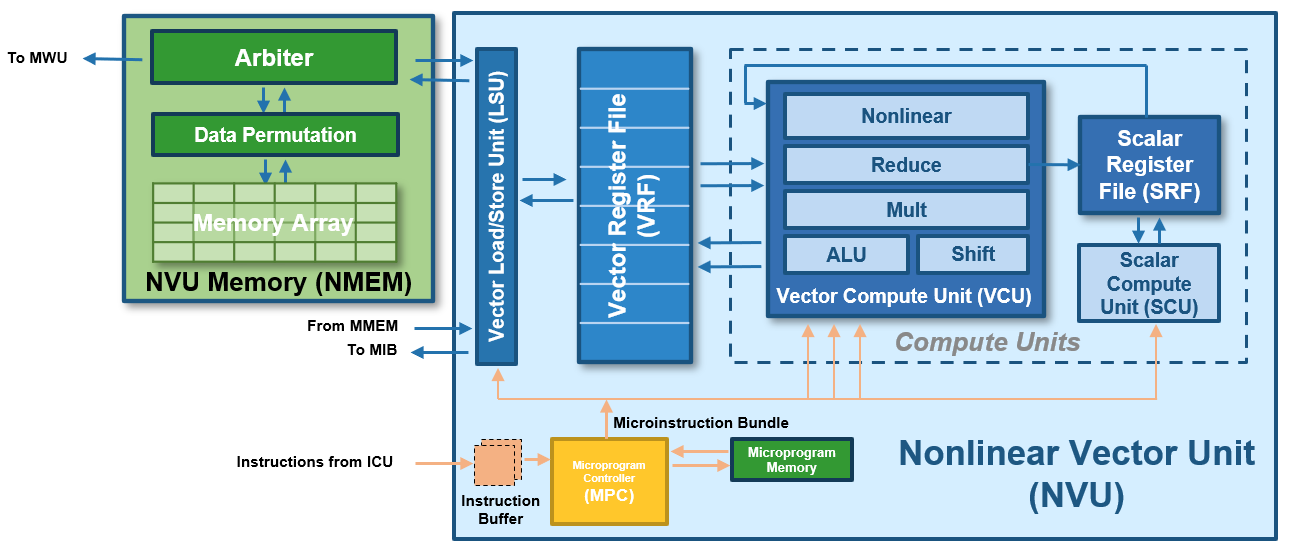}
	\centering
	\caption{Overall architecture of the NVU. The LSU transfers data between the VRF and either NMEM, MMEM, or the MIB. The vector compute unit (VCU) fetches operands from either the VRF or the SRF and can deposit results back to either the VRF or the SRF. The scalar compute unit fetches operands from the SRF and writes results back to the SRF. The microprogram controller (MPC) controls the operations of the NVU on a cycle-by-cycle basis.}
	\label{npe_architecture}
\end{figure*}

\subsection{Matrix Multiply Quantization}

The matrix multiply for transformer networks can be quantized to 16-bit fixed point with no perceptible loss in accuracy. Several works \cite{bhandare2019efficient, zafrir2019q8bert} have also shown the feasibility of 8-bit quantization of transformers. As shown in \cite{zafrir2019q8bert}, BERT can be implemented with 8-bit matrix multiplies with minimal accuracy loss. For this reason, while we support both 16 and 8-bit matrix multiplies, we plan to use 8-bit matrix multiply for our implementation. For both the 16-bit and the 8-bit matrix multiplies, the output of the MMU is written out to the MMEM (MMU scratchpad memory) as 16-bit fixed point values. Consequently, the NVU always consumes 16-bit fixed point values and generates either 8-bit or 16-bit results for the subsequent matrix multiplies.

\subsection{Software Simulation}
 
We simulate the architecture in software to validate the accuracy constraints of end-to-end BERT model inference. In order to model the overall accuracy loss, our simulations take into account the NPE modules used as well as the data quantization at each intermediate step. In particular, we model the fixed-point quantization effects of matrix multiplication and the approximation errors from our unified nonlinear processing approach, including piecewise linear approximations for various nonlinear operations.

\section{Nonlinear Vector Unit (NVU) Architecture} \label{NPE Architecture and Micro-architecture}

The key novel component of NPE's design is the nonlinear vector unit (NVU) that handles high-throughput nonlinear function computation with minimal resource overhead. The NVU is a data-parallel vector load/store architecture that performs arithmetic and logical operations on multiple elements in each clock cycle. The arithmetic performance of the NVU is defined by the width of the vector registers and the degree of parallelism in the arithmetic datapath.

The NVU, shown in Figure \ref{npe_architecture}, comprises a vector load/store unit (LSU), a vector register file (VRF), a vector compute unit (VCU), a scalar register file (SRF), a scalar compute unit (SCU), and a microprogram controller (MPC). The VRF, the VCU, and the LSU all operate on fixed-length vectors. The LSU accesses vectors resident in either the NVU memory (NMEM), the MMU memory (MMEM), or the MMU input buffer (MIB). The vector compute unit (VCU) and scalar compute unit (SCU) are designed to handle 16, 32 and 64-bit operations to allow for higher bit precision in intermediate calculations when required.

\subsection{Microprogram Controller (MPC)}

The instruction control unit (ICU) of NPE sends instructions to all NPE functional units, including the MMU and the NVU. These instructions are designed to operate over many clock cycles. The NVU interprets these high-level instructions to implement nonlinear functions like softmax, layer normalization, GELU, etc.

The microprogram controller (MPC) of the NVU is responsible for breaking down ICU instructions into a sequence of VLIW instruction bundles, which are used to control the LSU, the VCU, and the SCU. Since the VCU can execute up to three operations concurrently, as many as five instructions can be dispatched by the MPC in each micro-instruction cycle. The microprogram corresponding to each instruction is stored in the microprogram memory.

\subsection{Vector Register File (VRF)}

The vector register file (VRF) provides a high-throughput scratchpad for the vector compute unit (VCU). It allows the VCU to exploit locality of data across operations. The NVU requires a register file with 8 logical read and write ports in order to maintain concurrency of various VCU functional units. However, we implement the VRF using dual-port (1R1W) BRAMS with a combination of time sharing and data duplication. The resulting register file operates at the desired frequency of 200 MHz while requiring less than 5\% of overall BRAM resources.

\subsection{NVU Memory (NMEM)}

The NVU memory subsystem (NMEM) serves primarily as scratchpad for the NVU. The memory write unit (MWU) can also read results from the NMEM through a dedicated read port. The NMEM uses multiple banks of single-port memories, which are designed to support loads and stores of entire vectors in a single clock cycle. The NMEM can be implemented efficiently using single-port BRAMs on the FPGA. The NMEM also has arbitration logic to allow access from both the NVU and the MWU. The NMEM also contains logic for data permutation, which is required for strided and indexed accesses.

\subsection{Vector Load/Store Unit (LSU)}

The vector load/store unit performs fixed-length vector transfers between the vector register file and either the NMEM, the MMEM, or the MMU input buffer (MIB). The LSU supports non-strided, strided, and indexed loads and stores from the NMEM.

\subsection{Vector Compute Unit (VCU)}

The vector compute unit (VCU) supports a full complement of vector and intra-vector arithmetic (add, subtract, multiply, shift, sum, dot product, etc.), logical, compare ($<$, $\geq$, $=$, etc.), min/max and permute operations. In addition, specialized capabilities for piecewise polynomial approximation have been added, allowing the NVU to approximate nonlinear functions more than $10\times$ faster than traditional SIMD processors. The VCU reads vectors from the VRF or the scalar register file (SRF) and, depending on the operation, writes results back to the VRF or SRF. Vector reduction results are written to the SRF, and vector-scalar operations fetch scalar operands from the SRF.

The VCU implements multi-precision arithmetic while sharing resources between 8, 16, 32, and 64-bit fixed point data types. While the MMU only needs to handle a single data type (either 8 or 16-bit fixed point), NVU operations typically involve mixed precision including 32 and 64-bit representations for intermediate calculations.

The VCU can execute up to three operations concurrently.

\subsection{Scalar Compute Unit (SCU)}

The NVU also includes a scalar compute unit (SCU), which operates out of a scalar register file (SRF). Like the VCU, the SCU can handle 8, 16, 32, and 64-bit operations. The concurrent operation of vector and scalar functional units allow for the computation of nonlinear functions of vector reduce operations. For example, the NVU is capable of performing an inner product followed by the $\frac{1}{\sqrt{x}}$ operation for layer normalization variance calculations while maintaining full throughput.

\subsection{Scalable Architecture of the NVU}

All operations of the NVU are performed on fixed bit-width vector registers which make up the VRF. There are 32 vector registers in the VRF. All NVU micro-instructions reference source and destination vector registers. The overall performance of the NVU can be described by a single parameter, i.e. the vector register width ($VRWIDTH$). The number of elements processed per micro-instruction depends on the element size (8, 16, 32, or 64 bits). For example, a $VRWIDTH$ of 256 can hold 32 elements of 8 bits, 16 elements of 16 bits, etc. The NVU's area and performance depend on the number of elements that can be processed per micro-instruction.

\section{Throughput Analysis} \label{Experiments}

For initial analysis of NPE's architecture, we examine the throughput requirements for BERT on our architecture. While the MMU has both 8-bit and 16-bit variations, we focus on NPE with 16-bit MMU and pair it with NVUs of different $VRWIDTH$. For the remainder of this work, we describe the NVU variants based on the vector register width $VRWIDTH$. NVU-$VRWIDTH$ refers to the NVU with $VRWIDTH$-bit vector registers. For instance, NVU-256 means that a vector register is 256 bits. We focus on four NVU sizes (NVU-256, NVU-512, NVU-1024, and NVU-2048), comparing the throughput of each NVU variant to the required BERT throughputs.

\subsection{NVU Throughput}

Table \ref{NPE_cycles} shows the individual NVU performance results on each nonlinear function required for BERT. To normalize the results across NVU vector register widths, we give the number of cycles needed to process a 512 length array of 16-bit elements and the corresponding throughput in elements per cycle.

\begin{table}[!thb]
	\centering
	\caption{Throughput (elements per cycle) of NVU with different $VRWIDTH$ on BERT's nonlinear functions; cycle count to process a 512-element vector is shown in parentheses.}
	\resizebox{0.35\textwidth}{!}{%
		\begin{tabular}{@{}llll@{}}
			\toprule
			NVU Width & Softmax & Layer Norm & GELU \\
			\midrule
			NVU-256 & 1.64 \hspace{4pt}\textit{(312)} &  0.64 \hspace{4pt}\textit{(804)} & 4 \hspace{8pt}\textit{(128)} \\
			NVU-512 & 3.05 \hspace{4pt}\textit{(168)} &  1.29 \hspace{4pt}\textit{(396)} & 8 \hspace{8pt}\textit{(64)} \\
			NVU-1024 & 4.74 \hspace{4pt}\textit{(108)} & 2.42 \hspace{4pt}\textit{(212)} & 16 \hspace{4pt}\textit{(32)} \\
			NVU-2048 & 6.40 \hspace{4pt}\textit{(80)} & 4.13 \hspace{4pt}\textit{(124)} & 32 \hspace{4pt}\textit{(16)} \\
			\bottomrule
		\end{tabular}
	}
	\label{NPE_cycles}
\end{table}

\subsection{BERT Throughput Requirements}

We analyze the effective throughput requirement for each nonlinearity in BERT. This builds on the analysis in Table \ref{throughput}, where we established that the worst-case throughput requirement for softmax is 32 elements per cycle to keep up with the MMU. However, here we demonstrate that we can relax the worst case requirement by taking into account the optimization of overlapping independent computations. Then, we show the final requirements after optimization.

\subsubsection{\textbf{Overlapping Computation}}

\label{overlap}

In most cases, each stage of the transformer network computation is dependent on the results of the preceding stage. For instance, the feed-forward layer has a matrix multiply with GELU followed by a matrix multiply with Layer Normalization. The GELU computation must be finished before the next matrix multiply is started. This holds true for all Layer Normalization and GELU operations in BERT. This means that GELU and Layer Normalization must be rate matched with the MMU in order to avoid stalling the MMU.

Fortunately, this is not the case with softmax and parts of the attention mechanism. We can reduce the throughput requirements of softmax by overlapping it with independent matrix computations in multi-headed self-attention. For example, the computation in Table \ref{bert_compute} of softmax($(Q_i{K_i}^T)/k$) can be overlapped with the matrix multiplication $V_i$ = $X$ $W_{V_i}$. Since computation for each attention head is independent, we can also overlap softmax for head $i$ with some part of the computation for the next head $i+1$. In this way,the throughput requirements of the softmax computation can be relaxed by more than $4\times$.

\subsubsection{\textbf{Optimized BERT Throughput Requirements}}

Taking overlapping computations into account, we see the throughput requirements shown in Table \ref{throughput_opt}. In general, the throughput requirements of matrix multiplies in BERT do not depend on BERT network sequence length. However, for some of the attention computation, there is a dependence on sequence length. This only affects nonlinearity throughput when we overlap independent matrix multiplies with softmax, which is why we see a throughput dependence for softmax on sequence length in Table \ref{throughput_opt}.

Although softmax tends to have very high throughput requirements for higher sequence lengths, it only accounts for a small percentage of overall computation (see Table \ref{throughput}). Layer Normalization and GELU are needed for approximately two thirds of the total computation time but have relatively lower throughput requirements. If the NVU's softmax computation cannot match MMU throughput, we may still only get a small inference time overhead. Meanwhile, if layer normalization or GELU cannot be throughput-matched, we would expect a more noticeable inference time overhead.

\begin{table}[!thb]
	\centering
	\caption{Throughput  (elements per cycle) required for different BERT input sequence lengths on nonlinear functions for NPE (16-bit MMU); Layer Norm A refers to normalization after attention and Layer Norm B refers to normalization after GELU.}
	\resizebox{0.5\textwidth}{!}{%
		\begin{tabular}{@{}lllll@{}}
			\toprule
			Sequence Length & Softmax & Layer Norm A & Layer Norm B & GELU \\
			\midrule
			64 & 0.92 & 2.6 & 0.6 & 2.6 \\
			128 & 1.79 & 2.6 & 0.6 & 2.6 \\
			256 & 3.39 & 2.6 & 0.6 & 2.6 \\
			512 & 6.29 & 2.6 & 0.6 & 2.6 \\
			\bottomrule
		\end{tabular}
	}
	\label{throughput_opt}
\end{table}

By comparing results from Tables \ref{NPE_cycles} and \ref{throughput_opt}, we see that NVU-2048 is more than capable of keeping up with the 16-bit MMU. In fact, NVU-1024 can approach or exceed most of the requirements except softmax with sequence length 512. Given that softmax only takes up a few percentage of computation, and a sequence length of 512 is not needed for most applications, it is evident that there would only be marginal benefit of using NVU-2048 over NVU-1024. For this reason, we only analyze NVU-2048 for its inference time, as a comparison metric indicating ideal NVU performance (where MMU never stalls). Similarly, with the 8-bit MMU, the NVU-2048 also nearly matches the matrix multiplier throughput and can be used as a reference point.

\section{Evaluation} \label{Results}

We implement NPE at 200 MHz on the Xilinx Zynq Z-7100 FPGA, which has 2,020 DSP slices, 26.5 Mb RAM, and 277k LUTs. We look at several NVU variants (NVU-256, NVU-512, and NVU-1024), each of which can be paired with either the 8-bit or 16-bit MMU. We examine FPGA utilization for each NVU variant separately, then show overall FPGA utilization of each of the six resulting NPE configurations. We calculate software-simulated inference times for BERT for these six configurations and compare them to the corresponding NVU-2048 reference inference time. Finally, we evaluate NPE's performance on BERT inference relative to other implementations'.

\subsection{FPGA Utilization}

\begin{table*}[!thb]
	\centering
	\caption{FPGA Resource Utilization for components of NVU-256, NVU-512, and NVU-1024 on Zynq Z-7100.}
	\resizebox{0.7\textwidth}{!}{%
		\begin{tabular}{@{}ll|llllll@{}}
			\toprule
			Module & $VRWIDTH$ & LUT & FF & DSP Slices & BRAM & F7 Mux & F8 Mux\\
			\toprule
			
			NMEM & NVU-256 & 776 \hspace{2pt} \textit{(0.28\%)} & 1234 \textit{(0.22\%)} & 0 & 4 \hspace{2pt} \textit{(0.53\%)} & 0 & 0\\

			VRF & NVU-256 & 156 \textit{(0.06\%)} & 513 \hspace{2pt} \textit{(0.09\%)} & 0 & 4 \hspace{2pt} \textit{(0.53\%)} & 0 & 0 \\
			
			VCU+SCU & NVU-256 & 10328 \textit{(3.72\%)} & 1753 \textit{(0.32\%)} & 8 \hspace{2pt} \textit{(0.4\%)} & 0 & 3 \textit{(<0.01\%)} & 0 \\
			
			\midrule
			
			Total & NVU-256 & \textbf{11260} \textbf{(4.06\%)} & \textbf{3500} \textbf{(0.63\%)} & \textbf{8} \hspace{2pt} \textbf{(0.4\%)} & \textbf{8} \hspace{2pt} \textbf{(1.06\%)} & \textbf{3} \textbf{(<0.01\%)} & \textbf{0} \\

			\toprule
			
			NMEM & NVU-512 & 1330 \textit{(0.48\%)} & 2268 \textit{(0.41\%)} & 0 & 8 \hspace{2pt} \textit{(1.06\%)} & 0 & 0 \\
			
			VRF & NVU-512 & 306 \textit{(0.11\%)} & 1025 \textit{(0.18\%)} & 0 & 8 \hspace{2pt} \textit{(1.06\%)} & 0 & 0 \\
			
			VCU+SCU & NVU-512 & 19549 \textit{(7.05\%)} & 3441 \textit{(0.62\%)} & 16 \textit{(0.79\%)} & 0 & 12 \textit{(<0.01\%)} & 5 \textit{(<0.01\%)}\\
			
			\midrule
			
			Total & NVU-512 & \textbf{21185} \textbf{(7.64\%)} & \textbf{6734} \textbf{(1.21\%)} & \textbf{16} \textbf{(0.79\%)} & \textbf{16} \textbf{(2.1\%)} & \textbf{12} \textbf{(<0.01\%)} & \textbf{5} \textbf{(<0.01\%)} \\

			\toprule

			NMEM & NVU-1024 & 2902 \textit{(1.05\%)} & 4377 \textit{(0.79\%)} & 0 & 16 \textit{(2.1\%)} & 350 \textit{(0.25\%)} & 0 \\
			
			VRF & NVU-1024 & 607 \textit{(0.22\%)} & 2049 \textit{(0.37\%)} & 0 & 16 \textit{(2.1\%)} & 0 & 0 \\
			
			VCU/SCU & NVU-1024 & 34423 \textit{(12.41\%)} & 6984 \textit{(1.26\%)} & 32 \textit{(1.58\%)} & 0 & 37 \textit{(0.03\%)} & 5 \textit{(<0.01\%)} \\		
			
			\midrule
			
			Total & NVU-1024 & \textbf{37932} \textbf{(13.67\%)} & \textbf{13410} \textbf{(2.42\%)} & \textbf{32} \textbf{(1.58\%)} & \textbf{32} \textbf{(4.2\%)} & \textbf{387} \textbf{(0.28\%)} & \textbf{5} \textbf{(<0.01\%)} \\
	
			\bottomrule
		\end{tabular}
	}
	\label{mem_synth}
\end{table*}

\begin{figure}[b]
	\includegraphics[width=1\columnwidth]{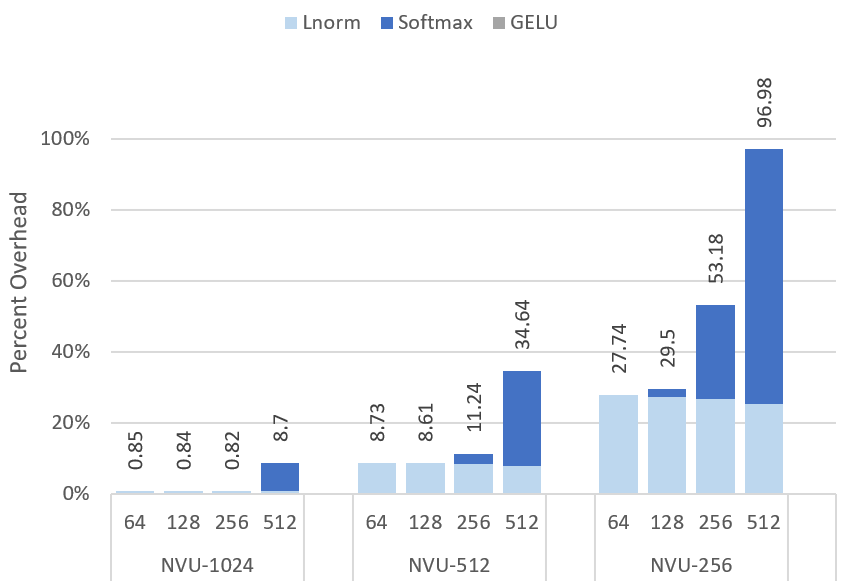}
	\centering
	\caption{BERT inference time percent overhead using NPE with 16-bit MMU for different sequence lengths and NVU variants. Overhead is relative to the minimum time using NVU-2048. }
	\label{npe_overhead}
\end{figure}

In Table \ref{mem_synth}, we individually show the FPGA utilization results for several components of the NVU: the NVU memory (NMEM), the vector register file (VRF), and the compute units (VCU and SCU). Then, in Table \ref{total_synth}, we give the cumulative FPGA resource utilization for NPE using each NVU variant, both for 8-bit and 16-bit NPE.

\begin{table}[t]
	\centering
	\caption{Overall FPGA Resource Utilization on Zynq Z-7100 for NPE with 8-bit and 16-bit MMU with NVU-256, NVU-512, and NVU-1024.}
	\resizebox{0.5\textwidth}{!}{%
		\begin{tabular}{@{}ll|lllll@{}}
			\toprule
			MMU & $VRWIDTH$ & LUT & FF & DSP Slices & BRAM \\
			\toprule
			
    			8-bit & NVU-256 & 165776 \textit{(59.76\%)} & 341151 \textit{(61.49\%)} & 1994 \textit{(98.71\%)} & 345  \textit{(45.70\%)} \\

			8-bit & NVU-512 & 175701 \textit{(63.33\%)} & 344385 \textit{(62.07\%)} & 2002 \textit{(99.10\%)} & 353  \textit{(46.75\%)} \\

			8-bit & NVU-1024 & 192448 \textit{(69.37\%)} & 351061 \textit{(63.28\%)} & 2018 \textit{(99.90\%)} & 369  \textit{(48.87\%)} \\
			
			\midrule

			16-bit & NVU-256 & 129231 \textit{(46.59\%)} & 250738 \textit{(45.19\%)} & 1995 \textit{(98.76\%)} & 502.5  \textit{(66.56\%)} \\

			16-bit & NVU-512 & 139156 \textit{(50.16\%)} & 253972 \textit{(45.78\%)} & 2003 \textit{(99.16\%)} & 510.5  \textit{(67.61\%)} \\

			16-bit & NVU-1024 & 155903 \textit{(56.20\%)} & 260648 \textit{(46.98\%)} & 2019 \textit{(99.95\%)} & 526.5  \textit{(69.73\%)} \\
			
			%\bottomrule
			
			%8 & OPU & 154516 \textit{(55.70\%)} & 337651 \textit{(60.86\%)} & 1986 \textit{(98.32\%)} & 337 \textit{(44.64\%)} \\	
            %16 & OPU &  117971 \textit{(42.53\%)} & 247238 \textit{(44.56\%)} & 1987 \textit{(98.37\%)} & 494.5 \\

			\bottomrule
		\end{tabular}
	}
	\label{total_synth}
\end{table}

From these results, we see that all the NVU variants are small relative to the overall NPE design. Even NVU-1024 uses less than three percent of overall flip-flop, DSP slice, and BRAM resources each. The larger NVUs do use 7-15\% of the overall LUT resources, much of which is due to the muxes required for shifting. Despite this, the overall design still has many LUTs left over.

\subsection{Inference Time}

The system simulation gives a cycle count estimate for a single inference of $\text{BERT}_{\text{BASE}}$, which can be used to determine inference time given the operating clock speed. The relative inference times of NPE with 16-bit MMU and NVU-256, NVU-512, and NVU-1024 are compared to inference time with NVU-2048. For NPE with 16-bit MMU, NVU-2048 gives the ideal inference time because it always exceeds the MMU throughput.

Figure \ref{npe_overhead} shows the percent inference time overhead of NVUs of different $VRWIDTH$ for NPE with 16-bit MMU. We see that in all cases GELU does not add latency overhead for any sequence length. Overall, NVU-1024 has very little overhead compared to the baseline case. The small difference is because layer normalization throughput is slightly lower than that which is needed to match the MMU. For smaller sequence lengths, NVU-1024 adds less than 1\% latency overhead, NVU-512 adds around 10\%, and NVU-256 adds about 30\%. Depending on the use case, these overheads may be acceptable given the reduced area costs. For higher sequence lengths, NVU-256 begins to show very large overheads of 53\% and 97\%.

Note that inference time overhead alone is not the only criteria that should be used to evaluate these options. Even larger overheads may be acceptable, as long as the overall inference time including overhead is within the target for conversational AI. For this reason, the actual inference time is compared below.

%From these results, it seems NVU-1024 and NVU-512 are both good options for most scenarios. NVU-256 adds too much overhead for most cases, but may be more useful for a smaller design with only 1024 multipliers. Note that with 8-bit MMU, we would see similar results 

\begin{figure*}[!th]
	\includegraphics[width=2\columnwidth]{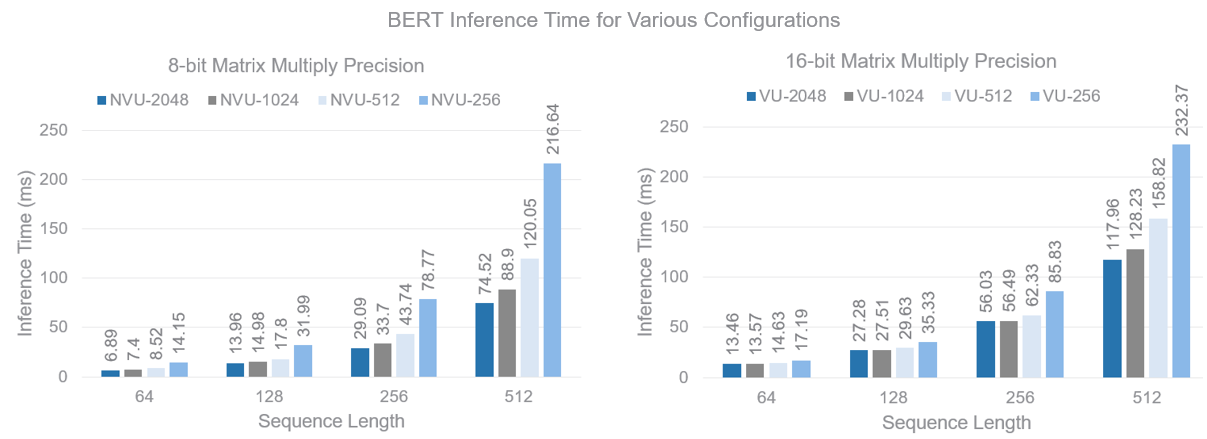}
	\centering
	\caption{BERT inference time (in ms) with different NVU widths and sequence lengths. Results are shown separately for NPEs with 8-bit and 16-bit matrix multiplies.}
	\label{npe_latency}
\end{figure*}

The BERT inference time for NPE with 16-bit and 8-bit MMUs with each $VRWIDTH$ is shown in Figure \ref{npe_latency}. We see that NPE with 8-bit MMU can achieve sub-10 ms inference time with sequence length of 64 even with NVU-512, but that the inference time increases proportionally as sequence length increases. For typical applications, sequence length of 64 is sufficient. For conversational AI, we require within 10-15 ms inference time, which we can clearly surpass with NVU-512 and NVU-1024 for both 8 and 16-bit.

\subsection{\textbf{Comparison with CPU, GPU, and FPGA}}

The authors of the FTRANS transformer FPGA accelerator \cite{li2020ftrans} provide inference benchmarks by running RoBERTa, an optimized version of BERT with the same model architecture but trained more thoroughly. Since BERT and RoBERTa have the same architecture, we can compare our BERT accelerator's inference times with their RoBERTa benchmarks. We compare with our NPE with 16-bit and 8-bit MMUs with NVU-1024 on the Zynq Z-7100. The devices used in the benchmark are an i7-8700k CPU, an RTX 5000 GPU, and an Ultrascale+ VCU118 FPGA (for FTRANS). The RTX 5000 has $1.52\times$ more compute units than our Zynq FPGA and runs at $8.1\times$ higher clock frequency. The VCU118 has 6,840 DSP slices and 2,586k logic cells ($3.39\times$ the DSP slices and $5.82\times$ the logic cells on our board). The inference times and relative latencies are shown in Table \ref{ftrans_comparison}. We also give the approximate power consumption of each device.

\begin{table}[!thb]
	\centering
	\caption{Throughput (inference/sec) of NPE with NVU-1024 compared with CPU (i7-8700k), GPU (RTX 5000), and FPGA (VCU118). We also give relative throughput compared to FTRANS, throughput per DSP slice relative to FTRANS (for FPGA implementations), and approximate power consumption.}
	\resizebox{0.48\textwidth}{!}{%
		\begin{tabular}{@{}l|llllll@{}}
			\toprule
			 & i7-8700k & RTX 5000 & FTRANS & NPE (16-bit) & NPE (8-bit)\\ % Jetson TX2 
			\midrule

			Throughput & 3.76 & 57.46 & 101.79 & \textbf{73.69} & \textbf{135.14} \\	 % 9.75	
			Relative Speedup & $0.037\times$ & $0.56\times$ & $1\times$ (baseline) & \textbf{0.72$\times$} & \textbf{1.33$\times$} \\ % 0.096x
			
			DSP Slices Utilized & - & - & 6,840 & \textbf{2,020} & \textbf{2,020} \\
			
			Throughput per DSP & - & - & 0.0148 ($1\times$) & \textbf{0.0365 ($2.5\times$)} & \textbf{0.0669 ($4.5\times$)} \\
			
			Approximate Power (W) & 80 & 120 & 25 & \textbf{20} & \textbf{20} \\
			\bottomrule
		\end{tabular}
	}
	\label{ftrans_comparison}
\end{table}

From the results, we see that the CPU is far too slow for conversational AI. While the RTX 5000 GPU gets close, it does not meet the conversational AI latency targets. However, with a larger or more optimized GPU it could meet these requirements, albeit with much higher power consumption. Both FTRANS and NPE implementations stay within the range needed for conversational AI.

\subsection{Benchmarks Discussion}

The biggest benefit of an FPGA implementation of BERT over CPU and GPU is with power consumption. From Table \ref{ftrans_comparison}, we see about a $4\times$ power benefit over CPU and $6\times$ over GPU. This difference in power consumption is especially important for NLP processing on edge devices. While FTRANS and NPE both have comparable performance and power, FTRANS uses over $3\times$ more resources than NPE since it uses a much larger FPGA. We attribute some of the difference in resource consumption to the fact that FTRANS uses specialized modules for each transformer and each nonlinearity, which leads to additional area and under-utilized components.

\section{Conclusion} \label{Conclusion}

In this paper we propose NPE, an FPGA-based overlay processor that is domain-specialized for Natural Language Processing. NPE offers software-like programmability and provides a unified framework to process arbitrarily complex nonlinear functions. If a new state-of-the-art NLP model were to surpass transformers in the coming years, NPE is most likely flexible enough to adapt to it without requiring reconfiguring the FPGA accelerator or adding specialized processing modules. NPE can also meet the inference latency requirements for conversational AI for the BERT language model. Relative to CPU and GPU, NPE has $4\times$ and $6\times$ lower power consumption respectively. Our accelerator shows comparable performance to a transformer model specialized FPGA accelerator, but NPE uses $3\times$ lower FPGA resources. Overall, we find that NPE is a promising solution for low-cost and low-power NLP network inference at the edge.

\bibliographystyle{ACM-Reference-Format}
\bibliography{sample}

\appendix

\end{document}